\documentstyle[12pt,a4]{article}
\newcommand{\be}{\begin{equation}}
\newcommand{\ee}{\end{equation}}
\begin{document}
\vspace{20.0mm}
\title{Emergence of Approximate Translation Invariance in Finite
Intervals as a Speed Selection Mechanism for Propagating Fronts}

\author{{ Stavros Theodorakis$^{a}$ and Epameinondas
Leontidis$^{b}$}\\
        $^{a}$ {\normalsize \it Physics Department, University
of Cyprus,
        P.O. Box 20537, Nicosia 1678, Cyprus}\\
$^b$ {\normalsize \it Chemistry Department, University of Cyprus,
P.O. Box
20537, Nicosia 1678, Cyprus}}

\maketitle

\vspace{35.0mm}
\begin{abstract}

\noindent
We introduce a new velocity selection criterion for fronts
propagating into unstable and metastable states.
We restrict these fronts to large $finite$ intervals in the
comoving
frame of reference and require that their
centers be insensitive to the locations of the ends of the finite
intervals, exhibiting thus effectively an approximate translation
invariance. Only one monotonic front has this behavior and its
velocity is the one that is physically selected. We present
analytic results
in the case of piecewise parabolic potentials,
and numerical results in other cases.

\end{abstract}

\bigskip
\begin{center}
\end{center}
\newpage
\section{The selection principle}

In many systems rendered suddenly unstable, propagating fronts 
appear. The determination of the speed of front propagation
into an unstable state has attracted attention recently,
since it cannot be achieved by simply solving the differential
equation in the comoving frame of reference on a
one-dimensional infinite domain. Indeed, there are many
such solutions on such a domain, even though the
propagating front in
practice always relaxes to a unique shape and velocity. Selection
principles have thus been formulated
to determine the observable front, without having
to solve directly the partial differential equation of motion for
a range of initial conditions. These principles have involved
concepts of linear and nonlinear marginal stability, of
structural stability and of causality[1], and all of them try to
deal with the puzzle of the reduction of the multiple solutions
to
the single observed one. All these selection principles examine
the
wave from the viewpoint of the moving front, the corresponding
wave equation being reduced then to an ordinary differential
equation involving the speed $v$ of propagation. This speed has
a unique value if the front is invading a metastable state,
but not so for the case of invasion into an unstable state. The
latter case has consequently attracted most of the attention.

\vspace*{3mm}
The selection principles mentioned above have been based on the
study of Fisher's dimensionless equation,
$u_{t}=$$u_{xx}$$+f(u)$, on the
interval $(-\infty,\infty)$ with $f(0)=f(1)=0$, the states $u=0$
and $u=1$ being unstable and stable fixed points respectively.
Aronson and Weinberger[2] have shown that sufficiently localized
initial conditions will evolve into an observable front
propagating with speed $v^{*}$, provided $f>0$ on (0,1).
This selected speed $v^{*}$ is the lowest speed for
which the above partial differential equation admits a
monotonic front joining the stable state
$u=1$ to the
unstable state $u=0$, and it satisfies $v^{*}\geq
2\sqrt{f^{'}(0)}$.
Any monotonic traveling wave solution $u(\xi)$ of Fisher's
equation,
with $\xi=x-vt$ being the coordinate in the comoving frame of
reference, is a propagating front with speed $v>0$ and
satisfies the
"steady state" ordinary differential equation
\be
\frac{d^{2}u}{d\xi^{2}}+v\frac{du}{d\xi}+f(u)=0,
\ee
with $u(-\infty)=1$, $u(\infty)=0$. Note that any speed $v>v^{*}$
will
give a monotonic front, though not the observed one. The
selection criteria seek to determine this lowest speed $v^{*}$,
without
solving the initial value problem, selecting thus one among the
multiple
possible fronts invading the unstable state. If the selected
speed is
determined by the linear order terms alone we have the pulled
case
$v^{*}=2\sqrt{f^{'}(0)}$ (linear marginal stability). If linear
analysis
fails, we have the pushed case $v^{*}>2\sqrt{f^{'}(0)}$
(nonlinear marginal stability). In the case of fronts
invading a metastable state though, only one front is possible
on
the interval
$(-\infty,\infty)$, and hence the selection is automatic[3].  
    
\vspace*{3mm}
We shall adopt here a unifying approach, showing that there
exists
a single selection principle at work for both the metastable and
unstable cases, and for the regimes of linear and nonlinear
marginal stability alike. This approach exploits the fact that
the observed front
is translationally invariant in the comoving frame of reference,
even on a large $finite$ interval,
in the sense that its location is effectively independent of the
ends of the interval.
We solve then the steady state equation of motion on a large
finite interval for an arbitrary given speed $v$, subject to the
appropriate boundary conditions, obtaining a certain solution.
This solution, however, will have approximate translational
symmetry, thus
becoming a physically observable front, only for a certain value
of $v$. It is
this value $v^{*}$ of $v$ that is experimentally observed. Thus
$the$ $selected$ $front$ $is$ $the$ $one$ $that$ $is$
$effectively$
$translationally$ $invariant$ $on$ $a$ $large$ $finite$
$interval$, in the comoving frame of reference.    

\vspace*{3mm}
Indeed, let us assume that $u(\xi)$ is the solution of Eq. (1)
subject to the boundary conditions $u(L_{1})=1$ and $u(L_{2})=0$,
with $L_{1}\ll L_{2}$. There is only
one such solution for a given velocity $v$. We define the
continuous potential $V(u)$, where $f(u)=-\partial V/\partial u$.
We multiply Eq. (1) with $du/d\xi$ and integrate from $L_{1}$ to
$L_{2}$, obtaining then
\be
v=\frac{V(0)-V(1)-\frac{1}{2}w^{2}(L_{2})
+\frac{1}{2}w^{2}(L_{1})}{\int_{L_{1}}^{L_{2}}w^{2}(\xi)d\xi},
\ee
with $w(\xi)=du/d\xi$. If $u(\xi)$ is going to be a physically
observable front on
this large, but finite, interval, it will have to be essentially
translationally
invariant. This means that $du/d\xi$ will be effectively zero in
the regions close to the
boundaries, $u$ having reached a fixed point there.
Consequently
$w(L_{1})$ and $w(L_{2})$ will tend to zero, while
$\int_{L_{1}}^{L_{2}}w^{2}(\xi)d\xi$
will be finite and practically independent of $L_{1}$ and
$L_{2}$,
as $L_{1}\rightarrow -\infty$ and $L_{2}\rightarrow\infty$. Hence
the
speed $v$ of Eq. (2) becomes independent of the endpoints of the
interval,
acquiring a unique value. In other words, only the front
with that particular speed can be translationally invariant.

\vspace*{3mm}
The requirement that the front be independent
of the ends of the finite interval selects therefore
the speed
\be
v^{*}=(V(0)-V(1))/\int_{L_{1}}^{L_{2}}w^{2}(\xi)d\xi,
\ee
with $L_{1}\rightarrow -\infty$ and $L_{2}\rightarrow \infty$,
as the speed of the physically
observed front. Note that no distinction has been made here
between
metastable and unstable states. Indeed, given $any$ particular
speed $v$,
we can find a front interpolating between the stable and the
unstable or metastable state, as we shall demonstrate below,
provided the
solution
is found on a finite interval. As the boundaries go
to
infinity, the value of the speed is restricted to $v^{*}$ and the
front becomes the one corresponding to the speed of Eq. (3). 
 
\section{Analytic examples}
We shall demonstrate the proposed selection principle through
analytic and numerical work, both for the unstable and
metastable cases. We shall adopt for our analytic work
the continuous potential
\begin{eqnarray}
V(u)&=&\alpha(1-4u^{2})/8\,\,\,\,\,\,\,\,\,\,\,if\,\,\,\,\,0\leq
u\leq 1/2,\nonumber\\
&=&\nu(1-2u)(3-2u)/8\,\,\,\,\,if\,\,\,\,\,u\geq 1/2,
\end{eqnarray}
where $\nu>0$ and $\alpha=1$ or $-1$, depending on whether the
state
$u=0$ is unstable or metastable, respectively. The corresponding
$f(u)$,
\begin{eqnarray}
f(u)&=&\alpha u\,\,\,\,\,\,\,\,\,\,\,if\,\,\,\,\,0\leq
u\leq 1/2,\nonumber\\
&=&\nu(1-u)\,\,\,\,\,if\,\,\,\,\,u\geq 1/2,
\end{eqnarray}
is piecewise linear and leads to exact analytic
solutions. A similar
piecewise linear model, but with $\xi$ in the interval
$(-\infty,\infty)$, has been used
for a discrete reaction-diffusion equation[4]. Piecewise
parabolic potentials, always on the interval $(-\infty,\infty)$,
have been used in nucleation and crystallization problems as
well[5].

\vspace*{3mm}
Eq. (1) will be solved subject
to the conditions $u(L_{1})=1$, $u(L_{2})=0$, where
$L_{1}\ll 0\ll L_{2}$. We shall assume that the midpoint of the
front
occurs at the point $\xi_{0}$, with $u(\xi_{0})=1/2$, noting that
both $u(\xi)$ and $du/d\xi$ have to be continuous at that point.
There are thus five boundary conditions that have to be
satisfied,
whereas the solution will involve five unknown parameters for any
given
value of $v$, namely $\xi_{0}$ and two constants for each linear
piece of $f(u)$. We expect therefore a unique solution for each
value
of $v$.

\vspace*{3mm}
Indeed, the exact solution of Eq. (1) for the $f$ of Eq. (5) is 
\begin{eqnarray}
u(\xi)&=&1-\frac{1}{2}e^{v(\xi_{0}-\xi)/2}\frac{\sinh[\sqrt{v^
{2}+4\nu}
(\xi-L_{1})/2]}{\sinh[\sqrt{v^{2}+4\nu}(\xi_{0}-L_{1})/2]}\,\,
\,\,\,if\,
\,\,\,L_{1}\leq\xi\leq\xi_{0}\nonumber\\
&=&\frac{1}{2}e^{v(\xi_{0}-\xi)/2}
\frac{\sinh[\sqrt{v^{2}-4\alpha}
(\xi-L_{2})/2]}{\sinh[\sqrt{v^{2}-4\alpha}(\xi_{0}-L_{2})/2]}
\,\,\,\,\,if\,\,\,\,\xi_{0}\leq\xi\leq L_{2},
\end{eqnarray}
where $\xi_{0}$ satisfies
\be
2v=\sqrt{v^{2}+4\nu}\coth[\sqrt{v^{2}+4\nu}(\xi_{0}-
L_{1})/2]
+\sqrt{v^{2}-4\alpha}\coth[\sqrt{v^{2}-4\alpha}(\xi_{0}-L_{2})
/2].
\ee
These relations hold even if $v^{2}\leq 4\alpha$. The solution
of Eq. (7)
gives $\xi_{0}$ as
a function
of the speed $v$. 

\vspace*{3mm}
${\bf (a)}$ ${\bf Unstable}$ ${\bf case:}$ ${\bf\alpha=1.}$

\vspace*{3mm}
(i) We examine the case $v>2$ first. If we require
$L_{1}\ll\xi_{0}\ll L_{2}$, then Eq. (7) reduces
to $\sqrt{v^{2}-4}=\sqrt{v^{2}+4\nu}-2v$. This equation has a
real solution, $v=v^{*}$$=(\nu+1)/\sqrt{2\nu-2}$,
provided $\nu\geq 3$. In other words,
for that particular value of $v$ the midpoint $\xi_{0}$ can be
anywhere in
the interval and cannot be determined, rendering thus the front
effectively translationally
invariant on the finite domain $[L_{1},L_{2}]$. The value $v^{*}$
is therefore the selected speed if $\nu\geq 3$. In fact,
the departures of $v$ from $v^{*}$ are miniscule and fall
exponentially
as the ends of the interval go to infinity.

\vspace*{3mm}
Since $v^{*}>2$ when
$\nu>3$,
we have a pushed case. If however $\nu<3$, Eq. (7) yields
\be
\xi_{0}\approx
L_{1}+2\frac{\coth^{-1}[(2v+\sqrt{v^{2}-4})/
\sqrt{v^{2}+4\nu}]}{\sqrt{v^{2}+4\nu}},   
\ee
and the front sticks to the left boundary.

\vspace*{3mm}
(ii) Consider the case $v<2$ next. Eq. (6) shows then that on the
interval $[\xi_{0},L_{2}]$ the solution
$u(\xi)$ is proportional to $\sin[\sqrt{4-v^{2}}(\xi-
L_{2})/2]$ and has multiple roots. Our front though
has to be monotonic, with no overshooting occurring. Thus
$u(\xi)$ must
become
zero only on the boundary $L_{2}$. Therefore the point $r=L_{2}$
$-2\pi/\sqrt{4-v^{2}}$, where the above sine becomes zero, must
lie outside the interval $[\xi_{0},L_{2}]$, and hence
$\xi_{0}>r$.
In that case Eq. (7) yields
\be
\xi_{0}=L_{2}-\frac{\pi}{\sqrt{4-v^{2}}}+\frac{2}{\sqrt{4-
v^{2}}}\tan^{-1}\Bigl[\frac{-
2v+\sqrt{v^{2}+4\nu}\coth[\sqrt{v^{2}+4\nu}(\xi_{0}-
L_{1})/2]}{\sqrt{4-v^{2}}}\Bigr].
\ee
This exact equation holds for $v<2$ and all values of $\nu$, and
determines
the location $\xi_{0}$ of the midpoint of the monotonic front.
Thus Eqs. (7) and (9) determine fully $\xi_{0}$ for a given $v$.
In fact, if $\nu<3$ and $v$ is just below 2, Eq. (9) yields
\be
\xi_{0}\approx L_{2}-2\pi/\sqrt{4-v^{2}}-(\sqrt{\nu+1}-2)^{-1},
\ee
or equivalently
\begin{eqnarray}
&&v\approx\sqrt{4-4\pi^{2}/[L_{2}-\xi_{0}+(2-\sqrt{\nu+1})^{-
1}]^{2}}.\nonumber
\end{eqnarray} 
Thus $v$ has a plateau as a function of $\xi_{0}$ at the value
$v=2$, if $\nu<3$.

\vspace*{3mm}
Fig. (1) shows $v$ as a function of the midpoint $\xi_{0}$ for
the cases
$\nu=9$ and $\nu=2$, for the potential of Eq. (4) with
$\alpha=1$.
For $\nu=9$,
the selected speed is $v^{*}=2.5$ (pushed case). We see that
for $v>2.5$ the front is located close to $L_{1}$, while it
shifts abruptly to $L_{2}$ when the speed becomes less than
$v^{*}$. When $v$ is equal to $v^{*}$, the graph has a plateau,
indicating that $\xi_{0}$ is pretty much undetermined, the
solution being thus effectively translationally invariant.
For the case $\nu=2$,
on the other hand, the front is located at $L_{1}$ when $v>2$,
as indicated by Eq. (8), while for $v$ just below 2 the location
of the front shifts abruptly to $L_{2}$, as indicated by Eq.
(10). We
see thus the appearance of a plateau at $v=2$, if $\nu=2$.
For that value of $v$ the front's location is rather
undetermined,
indicating that the front has acquired effectively a
translational invariance.
The selected speed is thus $v^{*}=2$ (pulled case).

\vspace*{3mm}
Our analytic example indicates then that for $\nu>3$ we have
the pushed case, the selected speed being $(\nu+1)/\sqrt{2\nu-
2}$, while for $\nu\leq 3$ we have the pulled case, the selected
speed being $v^{*}=2$. In both cases the selected speed
corresponds
to a plateau in the graph of $v$ versus $\xi_{0}$, due to the
emergence of an approximate translational invariance of the front
at that speed.
    
\vspace*{3mm}
${\bf (b)}$ ${\bf Metastable}$ ${\bf case:}$ ${\bf\alpha=-1.}$

\vspace*{3mm}
If $\nu>1$ the stable state is at $u=1$ and the metastable one
at $u=0$. The boundary conditions are once more $u(L_{1})=1$,
$u(L_{2})=0$, $u(\xi_{0})=1/2$, along with continuity of $u$ and
$du/d\xi$ at $\xi_{0}$. For a given value of $v$ there are then
five boundary conditions and five unknown parameters, two
constants for each linear piece plus $\xi_{0}$. There is thus a
single solution of Eq. (1) for any given value of $v$. This is
the solution given by Eqs. (6) and (7), but with $\alpha=-1$. We
can show that Eq. (7) reduces to $v=v^{*}=(\nu-1)/\sqrt{2\nu+2}$
if $L_{1}\ll\xi_{0}\ll L_{2}$. In other words, for that
particular value of $v$ the midpoint $\xi_{0}$ can be anywhere
in the interval, leaving the location of the front undetermined.
The front acquires thus an effective translational invariance at
that speed. This translational invariance is recognised in Fig.
(2a), where $\nu=7$, as a plateau in the graph of $v$ versus
$\xi_{0}$ at the speed $v^{*}=1.5$.

\vspace*{3mm}
Note that there is a solution to the metastable problem for any
$v$ on a finite domain in the comoving frame of reference, but
only for one $v$ on the infinite domain. Indeed, if the
metastable problem is solved on $(-\infty,\infty)$, then
$\xi_{0}$ cannot be determined due to the exact translational
symmetry, leaving us thus with five boundary conditions but only
four unknown
parameters, two for each linear piece. Consequently $v$ will also
have to be considered as a parameter to be determined, giving
thus a solution only for a unique value of $v$[3], which is
precisely the one selected by our selection principle. On a
semi-infinite domain, on the other hand, a continuum of values
is possible for $v$[6].

\vspace*{3mm}
In the
unstable case, exact translational symmetry on $(-\infty,\infty)$
reduces again the number of unknown parameters by one
($\xi_{0}$), but the boundary condition at $u=0$ is trivially
satisfied due to the existence of two decaying exponentials, and
thus we are left with four boundary conditions and four unknown
parameters, two for each linear piece, for a given value of $v$.
Solutions are thus possible for a continuum of values of $v$. 

\section{Numerical examples}
We can demonstrate our selection principle numerically as well,
for the case
\be
f(u)=u(b+u)(1-u)/b,
\ee
where for $0<b<1$ the states $u=-b$, $u=0$ and $u=1$ are
metastable,
unstable and stable, respectively. In fact, it was this
particular choice of $f(u)$ that was used when the concepts of
linear and nonlinear marginal
stability were first proposed[7]. That study found that for
$1>b>1/2$
the selected
speed for the front invading the unstable state is $v^{*}=2$,
while for $0<b<1/2$ it is $v^{*}=(2b+1)/\sqrt{2b}$. For the front
invading the metastable state the selected speed was found to
be $(1-b)/\sqrt{2b}$.

\vspace*{3mm}
We have solved Eq. (1) numerically on a finite $\xi$ domain for
the $f(u)$
of Eq. (11), with $b=1/8$, subject to the boundary conditions
$u(L_{1})=1$, $u(L_{2})=-b$ (metastable case). We found that the
plot
of $v$ versus the characteristic point $\xi_{0}$ of the front,
where
$\xi_{0}$ is
defined through the relation $u(\xi_{0})=1/2$, has a plateau at
$v=1.75$
(see Fig. 2b), indicating that at that speed the solution has
become approximately translationally invariant on the finite
$\xi$ domain. The speed on the plateau is precisely the one
selected by the marginal stability criterion[7].

\vspace*{3mm}
Furthermore, we have solved Eq. (1) numerically on a finite $\xi$
domain for
the $f(u)$ of Eq. (11), with $b=1$ and $b=1/8$, subject to the
boundary
conditions $u(L_{1})=1$, $u(L_{2})=0$ (unstable case). We find
again
that the plot of $v$ versus $\xi_{0}$, where $u(\xi_{0})=1/2$,
has a
plateau at $v=2$ and $v=2.5$, respectively (see Fig. 3),
indicating
that the solution acquires effectively translational invariance
there. These values are once again the ones known to be
selected[7].

\section{Concluding remarks}
We see then that requiring the solution to have approximate
translational invariance
on a finite interval in the comoving frame of reference results
in the selection of a speed for the front in
both the metastable and unstable cases. This speed is precisely
the one
given by marginal stability. A similar selection of a single
velocity
occurs when a cutoff is introduced, albeit on an infinite
domain[8]. The importance of the translational invariance has
also been noted in connection with the precursors of the
propagating fronts [9]. Indeed, the selected solution is the only
one with a legitimate translation mode in its stability spectrum.
Thus one way of understanding the marginal stability of the
selected solution is through requiring that the stability
operator of a physically realizable solution possess a
translation zero mode. 

\vspace*{3mm}
We can adopt then a selection principle that reads "the selected
front
is the one that is approximately translationally invariant
on a large finite interval, with
respect to the comoving frame of reference". This principle is
very
easy to implement for both the metastable and unstable cases,
especially
numerically. Indeed, it suffices to solve the comoving frame
equation on a large finite interval. For large speeds we expect
the midpoint of the
front to be close to the left boundary. As the speed $v$ is
lowered, the
midpoint suddenly moves to the right boundary. The speed $v^{*}$
at which
this sudden move occurs is the speed selected by the physically
observed
front.

\newpage
%

\newpage
\noindent
{\bf Figure Captions \hfill}

\begin{enumerate}

\item[\bf Figure 1:]
The speed $v$ as a function of the midpoint $\xi_{0}$
of the front invading the unstable state, for the $f(u)$ of Eq.
(5),
with $L_{1}=-15$ and $L_{2}=15$. The plateau is at $v=2.5$ for
$\nu=9$,
and at $v=2$ for $\nu=2$. All quantities are dimensionless.

\item[\bf Figure 2:]
(a) The speed $v$ as a function of the midpoint $\xi_{0}$
of the front invading the metastable state, for the $f(u)$ of
Eq. (5), with $\nu=7$, $L_{1}=-15$ and $L_{2}=15$.
The plateau is at $v=1.5$. (b) The speed $v$ as a function of the
point $\xi_{0}$ of the front invading the metastable state, for
the $f(u)$ of Eq. (11), with $b=1/8$, $L_{1}=-15$ and $L_{2}=15$.
The plateau is at $v=1.75$.

\item[\bf Figure 3:]
The speed $v$ as a function of the midpoint $\xi_{0}$
of the front invading the unstable state, for the $f(u)$
of Eq. (11), with $L_{1}=-15$ and $L_{2}=15$. The plateau is at
$v=2$
for $b=1$, and at $v=2.5$ for $b=1/8$.

\end{enumerate}

\begin{thebibliography}{99}
\bibitem{Wim89}
Wim van Saarloos, Phys. Rev. {\bf 39A}, 6367 (1989); G.C.
Paquette, L.-Y. Chen, N. Goldenfeld, Y. Oono, Phys. Rev. Lett.
{\bf 72}, 76 (1994); G.C. Paquette and Y. Oono,
Phys. Rev. {\bf 49E}, 2368 (1994); J.-M. Chomaz and A. Couairon,
Phys. Rev. Lett. {\bf 84}, 1910 (2000); X.Y. Wang, S. Fan and T.
Kyu, Phys. Rev. {\bf 56E}, R4931 (1997). 

\bibitem{Aro78}
D.G. Aronson and H.F. Weinberger, Adv. Math. {\bf 30}, 33 (1978).

\bibitem{Cro93}
M.C. Cross and P.C. Hohenberg, Rev. Mod. Phys. {\bf 65}, 851
(1993).

\bibitem{Fath98}
Gabor Fath, Physica {\bf 116D}, 176 (1998).

\bibitem{Gra2000}
L. Granasy and D.W. Oxtoby, J. Chem. Phys. {\bf 112}, 2399
(2000);
L. Granasy and D.W. Oxtoby, J. Chem. Phys. {\bf 112}, 2410
(2000); C.K. Bagdassarian and D.W. Oxtoby, J. Chem. Phys. {\bf
100}, 2139 (1994); D.W. Oxtoby and P.R. Harrowell, J. Chem. Phys.
{\bf 96}, 3834 (1992); P.R. Harrowell and D.W. Oxtoby, J. Chem.
Phys.
{\bf 86}, 2932 (1987); H. Lowen and D.W. Oxtoby, J. Chem. Phys.
{\bf 93},
674 (1990). 

\bibitem{Mei83}
T. Meister and H. Muller-Krumbhaar, Phys. Rev. Lett. {\bf 51},
1780 (1983).

\bibitem{Benj85}
E. Ben-Jacob, H. Brand, G. Dee, L. Kramer and J.S. Langer,
Physica
{\bf 14D}, 348 (1985).

\bibitem{Bru97}
E. Brunet and B. Derrida, Phys. Rev. {\bf 56E}, 2597 (1997).

\bibitem{Kes98}
David A. Kessler, Zvi Ner, Leonard Sander, Phys. Rev. {\bf 58E},
107 (1998).



\end{thebibliography}
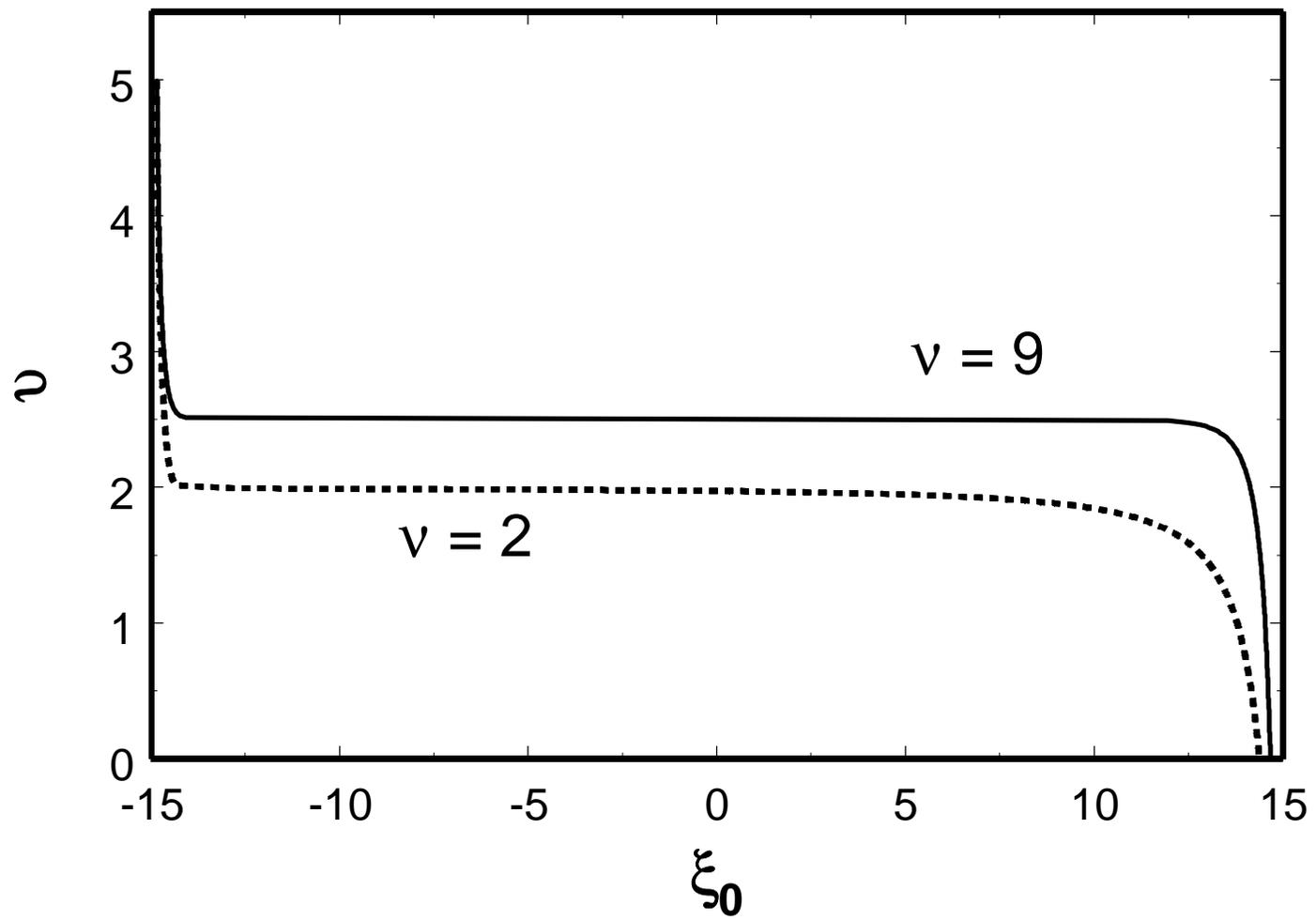
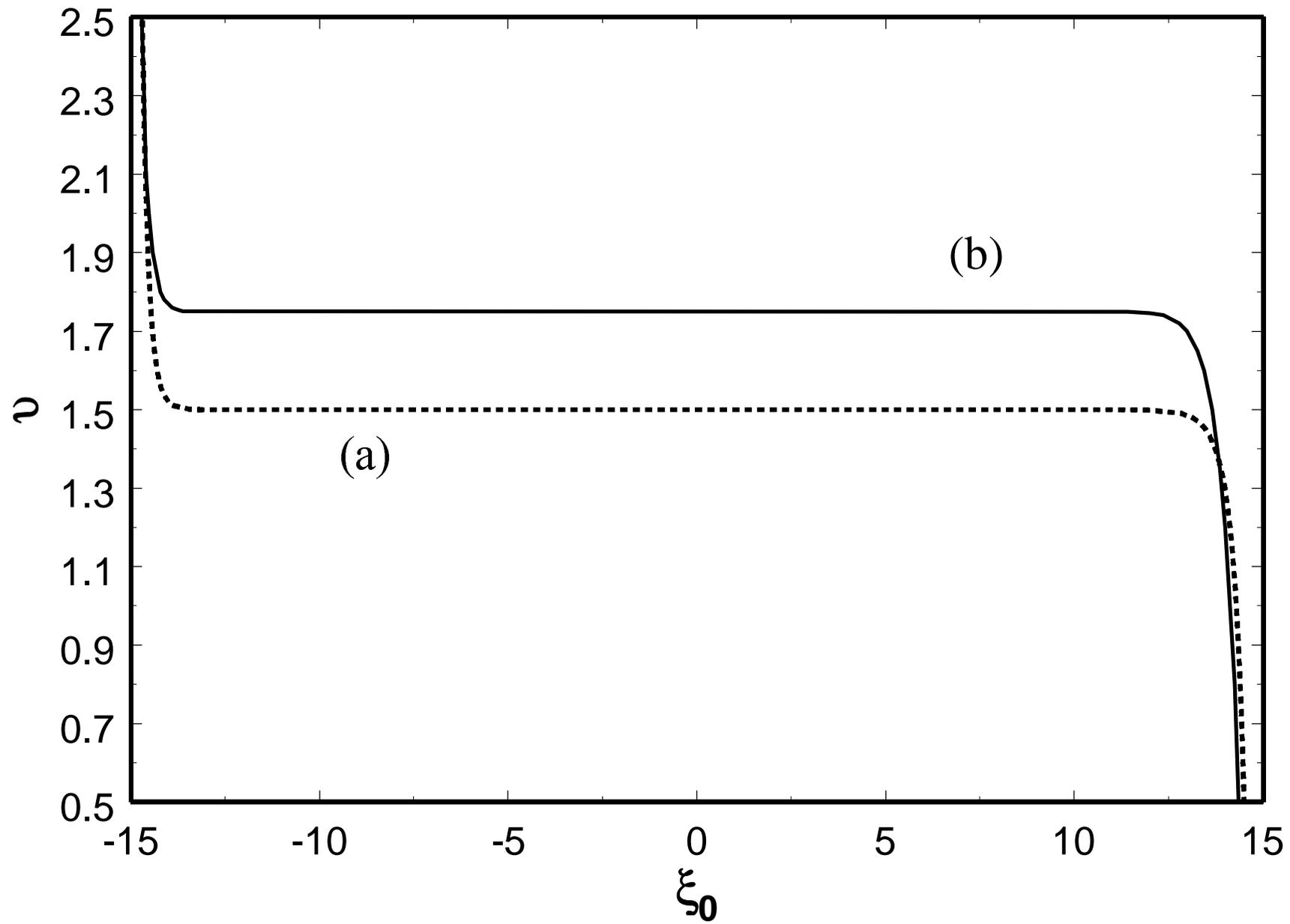
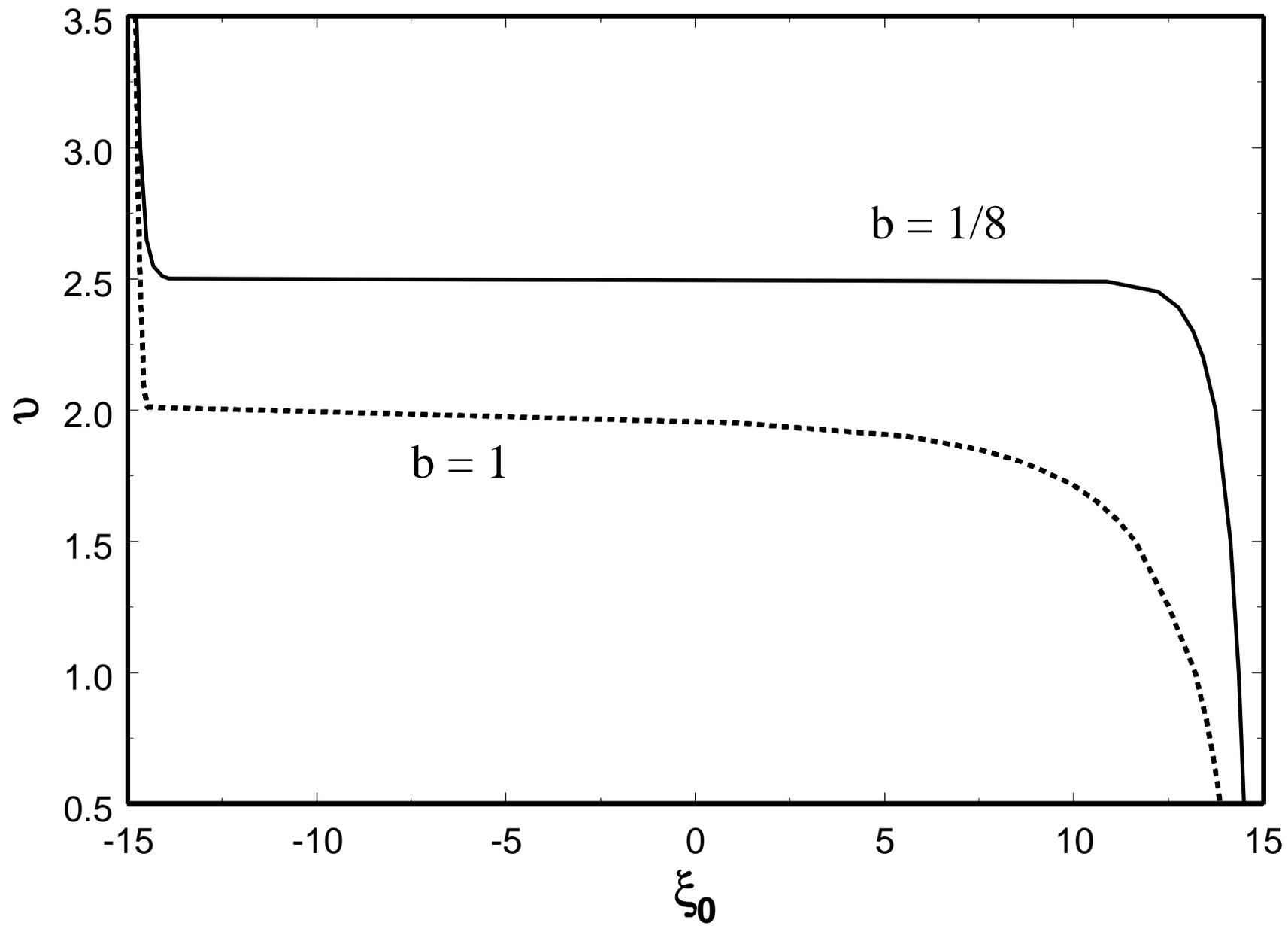
\end{document}